\documentclass[a4paper,11pt]{article}
\pdfoutput=1 % if your are submitting a pdflatex (i.e. if you have
\usepackage{jheppub} % for details on the use of the package, please
\usepackage[T1]{fontenc} % if needed
\usepackage{url}
\usepackage{setspace}
\usepackage{amsmath,amssymb,amscd, amsthm}
\usepackage{bbold}
\usepackage{amsfonts}
\usepackage{graphicx}
\usepackage{subcaption}
\usepackage{comment}

\usepackage{bm}

\numberwithin{equation}{section}

\gdef\ffrac#1#2{\textstyle\frac{#1}{#2}\displaystyle}
\gdef\be{\begin{equation}}
\gdef\ee{\end{equation}}
\gdef\e{\epsilon}
\gdef\a{\alpha}
\gdef\p{\partial}

\gdef\vZ{{\mathfrak Z}}

\gdef\zb{{\bar z}}
\gdef\d{{\delta}}

\title{\boldmath The $T\overline T$ deformation of quantum field theory as random geometry}

\author[a,b]{John Cardy}

\affiliation[a]{Department of Physics, University of California, Berkeley CA 94720, USA}
\affiliation[b]{All Souls College, Oxford OX1 4AL, UK}

\emailAdd{cardy@berkeley.edu}

\abstract
{We revisit the results of Zamolodchikov and others on the deformation of two-dimensional quantum field theory by the determinant $\det T$ of the stress tensor, commonly referred to as $T\overline T$. Infinitesimally this is equivalent to a random coordinate transformation, with a local action which is, however, a total derivative and therefore gives a contribution only from boundaries or nontrivial topology. We discuss in detail the examples of a torus, a finite cylinder, a disk and a more general simply connected domain.  
In all cases the partition function evolves according to a linear diffusion-type equation, and the deformation may be viewed as a kind of random walk in moduli space.  
 We also discuss possible generalizations  to higher dimensions.}

\begin{document} 
\maketitle
\flushbottom

\section{Introduction}\label{sec:intro}

In a remarkable unpublished paper in 2004 Zamolodchikov \cite{Zam} obtained simple analytic results for the expectation value of the operator 
$T_{zz}T_{\bar z\bar z}-T^2_{z\bar z}$ in any two-dimensional relativistic quantum field theory, where the $T_{\mu\nu}$ denote the components of the (euclidean) energy-momentum, or stress, tensor in complex coordinates. In particular he showed that it is well-defined by point-splitting, and that its vacuum expectation value is proportional to $-\langle T^\mu_\mu\rangle^2$. 

Furthermore he showed that, when the field theory is compactified on a spatial circle of circumference $R$, the expectation value of the above operator in any eigenstate of energy and momentum is simply related to the expectation values of $T^\mu_\nu$ itself. This leads to a simple differential equation for how the energy spectrum at finite $R$ evolves on deforming the action by a term $\propto T\overline T$. 

This has become known as the $T\overline T$ deformation of the original two-dimensional local field theory, and in recent years has become the object of some attention, one reason being that it turns out to be an example of a local field theory perturbed by irrelevant (non-renormalizable) operators which nevertheless has a sensible UV completion which is not, however, itself a local QFT. The fact that this occurs only for a special type of deformation has been termed asymptotic fragility, as opposed to asymptotic safety.  Dubovsky, Flauger and Gorbenko
\cite{Dub00} argued that an example corresponds to an exponential dressing of a free boson $S$-matrix by CDD factors which describes world-sheet scattering for the Nambu-Goto string, and Dubovsky, Gorbenko and Mirbabayi \cite{Dub0} gave an expression for such a dressing of the full $S$-matrix in the more general non-integrable case also. (That irrelevant operators should generate CDD factors was also pointed out by Mussardo and Simon \cite{Mus}.)
Dubovsky, Gorbenko, and Mirbabayi \cite{Dub} argued that the deformation corresponds to the dressing of the theory by Jackiw-Teitelboim \cite{Jac,Tei} gravity, a correspondence which has recently been demonstrated explicitly by the exact computation of the torus partition function by 
Dubovsky, Gorbenko, and Hern\'andez-Chifflet \cite{Dub2}.

The connection between this and the  $T\overline T$ deformation was established by Caselle, Fioravanti, Gliozzi and Tateo \cite{Cas}, who pointed out that the spectrum of states on a circle of a $T\overline T$ deformed CFT, first obtained in \cite{Zam}, when applied to a free boson is that of the Nambu-Goto string. 
Smirnov and Zamolodchikov \cite{Smi} also pointed out that the deformation corresponds to the  modification of an integrable $S$-matrix by a CDD factor, and extended these ideas to deformations built out of higher spin currents.  A different generalization involving a $U(1)$ current was studied by Guica \cite{Gui}.
Cavagli\`a, Negro, Sz\'ecs\'enyi and Tateo \cite{Cav} rederived and extended these results and verified them in the case of some integrable models. They also suggested a generalization of Zamolodchikov's results on the spectrum to the case open boundaries. 

McGough, Mezei and Verlinde \cite{Ver} studied the holographic interpretation (see also Refs.~\cite{Giv,Shy,Gir,Kra}). 
Cardy \cite{Car}, and McGough \em et al. \em\cite{Ver} considered the effect of the deformation on signal propagation in a non-trivial background. The hydrodynamics of adding such a term has been considered by Bernard and Doyon \cite{Ber}.

However, it has remained something of a puzzle as to why this particular deformation of a two-dimensional field theory is in some sense solvable, and to derive other results beyond those for the spectrum and $S$-matrix. In this paper we give a simple explanation of this fact, re-derive some of the known results and uncover some new ones. First, we point out that the infinitesimal deformation of the action is in fact proportional to the determinant $\det T$ of the stress tensor, which reduces to $T\overline T$ only in the limit of a conformal field theory (CFT). (However we shall continue to use the term `$T\overline T$ deformation', in line with the existing literature.) In two dimensions, $\det T$ is quadratic in $T_{ij}$, and this term in the action may therefore be decoupled by introducing an auxiliary field $h_{ij}$ coupled linearly to $T_{ij}$, itself with a local quadratic action also proportional to $\det h$. For an infinitesimal deformation this corresponds to an infinitesimal change in the metric $g_{ij}\to g_{ij}+h_{ij}$. Thus adding an infinitesimal term $\propto \det T$ may be thought of as coupling the theory to a random, locally correlated geometry. 

However, we show that (a) it is sufficient to restrict this to flat metrics, which correspond to infinitesimal diffeomorphisms $x_i\to x_i+\a_i(x)$, and (b)
the resulting quadratic action for $\a_i$ is then a \em total derivative. \em
It is this simple fact which is at the root of the solvability of the deformation. Since only $h_{ij}$, rather than $\a_i$, needs to be single-valued, it implies that  contributions to the total action for $h_{ij}$ can only occur if the domain $\cal D$ has non-trivial topology, as for the cylinder or torus, or if $\cal D$ has a boundary. In the cases where $\cal D$ has continuous spatial symmetries, such as the torus or disk, the measure on $h_{ij}$ then turns out to be concentrated on spatially uniform metrics, making the integration over them straightforward. 

This leads to linear differential equations for the evolution, parametrized by a coupling $t$, of the partition function $Z^{(t)}$ under the deformation. In fact it turns out that the equations satisfied by the partition function per unit area, $\vZ^{(t)}\equiv Z^{(t)}/A$ are simpler.\footnote{In an earlier version of this paper, this factor of the area was overlooked, as was the term $\propto-1/R$ in the case of the disk.}

As an example we quote (\ref{3.26b}) for a torus formed by identifying opposite edges of a parallelogram whose vertices lie at $\big((0,0),(L_1,L_2),\\ (L'_1,L'_2),(L_1+L'_1,L_2+L'_2)\big)$ :
 \be
\frac{\partial \vZ^{(t)}}{\partial t}=\left(\frac\p{\p L_1}\frac\p{\p L'_2}-\frac\p{\p L_2}\frac\p{\p L'_1}\right)
\vZ^{(t)}\,,
 \ee
 which may be expressed more compactly as
 \be\label{1}
 \p_t \vZ=(\p_L\wedge\p_{L'})\,\vZ\,.
 \ee
In particular, (\ref{1}) is  equivalent to known results for the deformation of the spectrum on a cylinder \cite{Zam,Smi}.

Similarly (\ref{3.48c}) for a finite cylinder  of circumference $L$ and length $L'$
 \be\label{2}
 \p_t\vZ=\p_L\p_{L'}\,\vZ\,,
\ee
where now $\vZ=Z/L$, and (\ref{4.18d}) for a disk of radius $R$,
\be\label{3c}
\p_t\widetilde Z=(1/4\pi)(\p_R-1/R)\p_{R}\,\widetilde Z\,,
\ee
(where the tilde indicates an average over gaussian shape fluctuations.)

Although equations like (\ref{1}, \ref{2}, \ref{3c}) are very simple, because of the non-trivial initial and boundary conditions they turn out to have interesting solutions. Since they are linear they may be solved by Green's function methods, but the choice of the integration domain and contour must be determined from other considerations. 

For the torus and finite cylinder, we expect that $Z^{(t)}$ may be written as a sum of terms of the form $e^{-LE^{(t)}(L')}$. In  these cases we then find that the functions $E^{(t)}$ satisfy the inviscid Burgers' equation (\ref{35}), as noted in the first case by Zamolodchikov \cite{Zam,Smi} and also Cavaglia \em et al.\em\ \cite{Cav} for the finite cylinder. These special solutions also appear to arise by solving the PDE using the heat kernel with a saddle point method, but in these cases the CFT initial conditions then allow the one-loop approximation to be exact. This has been noted for the torus in Ref.~\cite{Dub2}. For $t<0$ (in our convention, which corresponds to $\alpha<0$ in \cite{Smi} and $\mu<0$ in \cite{Ver}) all the solutions for $E^{(t)}$ remain regular up to a finite $|t|$ (or up to a finite $L'$ at fixed $t<0$), at which the ground state energy becomes singular, signaling what has been interpreted as a Hagedorn transition \cite{Cav}. However for $t>0$ and CFT initial conditions there is an infinite number of singular terms which condense at $t=0+$, so the expansion diverges for any finite $t>0$. Nevertheless, as we discuss in Sec.~\ref{sec3.3d}, the PDE for the partition function has a perfectly regular solution for $t\geq0$, which, however does not have a convergent spectral  decomposition.
It was also shown in Refs. \cite{Car,Ver} that $t,\alpha,\mu>0$ leads to superluminal signal propagation at finite energy density, so may not correspond to a sensible UV completion in Lorentzian signature. 

For the disk and other simply connected domains, the solutions exist but
in general have essential singularities as $t\to0+$, for example of the form $e^{-R^2/t}$ for the disk, as well as regular terms. Indeed, it may be checked that a solution as a formal power series expansion in $t$ is divergent. 
This suggests that the deformed theory with $t<0$  may have some difficulties in a general simply connected euclidean domain.
This may not be a serious impediment to its definition in Lorentzian signature, however.

Equations like (\ref{1},\ref{2}) may obviously be interpreted as arising from a stochastic process in the space of parameters of the domain. The deformed partition function in the original domain is the same as the undeformed partition function in the evolved domain, averaged over the stochastic process:
\be
Z^{(t)}({\cal D}_0)=\overline{Z^{(0)}({\cal D}_t)}\,.
\ee

 However the random walk interpretation is tricky as some of the noises turn out to be imaginary. This is related to the contour deformations necessary to define the integral over random metrics. Nevertheless one may argue that these domains  evolve towards a more symmetrical shape, as they, on average, shrink.

\section{Equivalence to a random metric problem}\label{sec2}
We consider a sequence of two-dimensional euclidean field theories ${\cal T}^{(t)}$, parametrized by a real parameter $t$. It is assumed that they each possess a local stress tensor $T^{(t)}_{ij}$ with the usual property of generating infinitesimal local changes in the metric. In this paper we consider only compact manifolds which may be endowed with a flat metric. If they are not simply connected, we dissect them into simply connected components $\cal D$.

For a theory to be translationally and rotationally invariant $T^{(t)}_{ij}$ is thus conserved and symmetric. ${\cal T}^{(0)}$ is assumed to be a conventional QFT with a UV limit corresponding to a CFT. To deform the theory infinitesimally from ${\cal T}^{(t)}$ to ${\cal T}^{(t+\delta t)}$ we formally add a term\footnote{The sign and factors of 2 are chosen so that our parameter $t$ matches the $\a$ of Refs.~\cite{Zam,Smi}, using our normalization of $T$.}
\be\label{eq:O}
-\delta t\,\,{\cal O}^{(t)}=-4\delta t\int \big(\det T^{(t)}\big)d^2x
\ee
to the action. Less formally, we define the deformation of a correlation function as 
\be
\delta\langle\Phi^{(t)}_1(x_1)\Phi^{(t)}_2(x_2)\cdots\rangle=
\delta t\,\langle{\cal O}^{(t)}\Phi^{(t)}_1(x_1)\Phi^{(t)}_2(x_2)\cdots\rangle^{(t)}_c\,,
\ee
and of  the free energy as
\be
\delta F^{(t)}=-\delta\log Z^{(t)}=-\delta t\,\langle{\cal O}^{(t)}\rangle_{{\cal T}^{(t)}}\,.
\ee
Note that $t$ has dimension (length)$^2$, so that the deformation is irrelevant in the IR, and conversely relevant in the UV. 

In Cartesian coordinates, $\det T=T_{11}T_{22}-T_{12}^2$. This is minus Zamolodchikov's operator \cite{Zam}
\be
T\overline T-\Theta^2=T_{zz}T_{\bar z\bar z}-T_{z\bar z}^2=\ffrac14\big[(T_{11}-T_{22}-2iT_{12})(T_{11}-T_{22}+2iT_{12})
-(T_{11}+T_{22})^2\big]\,,
\ee
 but  the expression
\be
\det T=\ffrac12\e_{ik}\e_{jl}T^{ij}T^{kl}
\ee
is more useful.

It is very important \cite{Smi} that the infinitesimal deformation is defined in terms of the stress tensor $T^{(t)}$ of the \em deformed \em theory. The result is not the same as adding a term $t\int\big(\det T^{(0)}\big)d^2x$ with a finite coupling $t$ to the action of ${\cal T}^{(0)}$.

The perturbation of the action (\ref{eq:O}), which is quadratic in $T^{(t)}$, may as usual be decoupled by a gaussian integral (Hubbard-Stratonovich transformation):
\be\label{9}
e^{2\delta t\int_{\cal D}\e_{ik}\e_{jl}T^{ij}T^{kl}d^2x}\propto \int[dh]e^{-(1/8\delta t)\int\int_{\cal D}\e^{ik}\e^{jl}h_{ij}h_{kl}d^2x
+\int_{\cal D}h_{ij}T^{ij}d^2x}\,,
\ee
where we have used $\e^{ik}\e^{jl}\e_{km}\e_{ln}=\delta^i_m\delta^j_n$, and suppressed the $t$-dependence of $T^{(t)}$. The integral is over a tensor field $h_{ij}$, of which the antisymmetric part decouples. As we shall only need to consider the saddle point solution and the relative gaussian fluctuations about this, we can afford to be cavalier about the precise integration contours.

Since it will turn out that\footnote{We may always assume this since we only ever consider infinitesimal $\delta t$, with a new $T^{(t)}_{ij}$ at each step.} $h_{ij}=O(\delta t)$, by the definition of $T_{ij}$ the second term is equivalent to an infinitesimal change in the metric $g_{ij}=\delta_{ij}+h_{ij}$. 
In principle the integration in (\ref{9}) is over all metrics infinitesimally close to euclidean, including those with non-zero curvature. However, it turns out that because $h_{ij}$ couples to a conserved tensor $T_{ij}$, this is not the case: we may restrict $h_{ij}$ to be an infinitesimal diffeomorphism
\be\label{2.7e}
h_{ij}=\a_{i,j}+\a_{j,i} \quad\mbox{with}\quad \a_{i,j}=\a_{j,i}\,,
\ee
that is, the metric is flat. 
It turns out then that the action for $\a$ is a total derivative and therefore receives contributions only from the boundary $\p\cal D$. This we now show.

Since the integral in (\ref{9}) is gaussian, its value is given by the value of the exponent at a saddle point
\be\label{2.8c}
h_{ij}=h_{ij}^*=(4\delta t)\e_{ik}\e_{jl}T^{kl}=(4\delta t)\big(g_{ij}T^k_k-T_{ij}\big)\,,
\ee
or equivalently
\be\label{2.9c}
T_{ij}=(1/4\delta t)\e_{ik}\e_{jl}{h^*}^{kl}=(1/4\delta t)\big(g_{ij}{h^*}^k_k-h^*_{ij}\big)\,.
\ee

In Cartesian coordinates this is simply
\be
T_{11}=(1/4\delta t)h^*_{22}\,,\quad T_{22}=(1/4\delta t)h^*_{12}\,,\quad
T_{12}=T_{21}=-(1/4\delta t)h^*_{21}=-(1/4\delta t)h^*_{12}\,,
\ee
so the conservation equations $T_{ij,j}=0$ imply
\be
h^*_{22,1} = h^*_{21,2}\,,\quad
h^*_{11,2} = h^*_{12,1}\,,
\ee
with solutions 
\begin{eqnarray}
&&h^*_{11} = 2\a_{1,1}\,,\quad h^*_{12} = 2\a_{1,2}\\
 &&h^*_{22} = 2\a_{2,2}\,,\quad h^*_{21} = 2\a_{2,1}\,,
 \end{eqnarray}
 where $\a_1$, $\a_2$ are arbitrary differentiable functions. 
 This is (\ref{2.7e}) since $h^*_{ij}$ is symmetric.

Now consider the quadratic part of the action at the saddle point\footnote{From now on we work in Cartesian coordinates unless explicitly stated and so lower all indices for convenience.}
\be
\e_{ik}\e_{jl}h^*_{ij}h^*_{kl}
=4\e_{ik}\e_{jl}(\p_i\a_j)(\p_k\a_l)=4\p_i(\e_{ik}\e_{jl}\a_j\p_k\a_l)\,.
\ee
Thus the total action, for a simply connected region, integrates up to a boundary contribution only
\be\label{23c}
 (1/2\delta t)\int (\e_{ik}\e_{jl}\a_j\p_k\a_l)dn_i-2\int \a_jT_{ij}dn_i\,,
 \ee
 where $n_i$ is the outward pointing normal.
 
 This can also be written
 \be\label{2.23d}
 (1/2\delta t)\int (\e_{jl}\a_j\p_k\a_l)ds_k-2\int \e_{ik}\a_jT_{ij}ds_k\,,
 \ee
 where $ds_k$ is the tangent vector.

  The above argument shows that
 the metric is flat and that the action is a total derivative 
 just at the saddle point. Because the action is quadratic, however, this is sufficient.
 The action has the form
\be
S=(1/2\delta t)h\cdot M\cdot h+h\cdot T
\ee
and the saddle is at 
\be
h=h^*[T]=-\delta t M^{-1}\cdot T
\ee
If we write $h=h^*[T]+\tilde h$,
\be
S=S[h^*[T],T]+(1/2)\tilde h\cdot M\cdot \tilde h
\ee
then the integral over $\tilde h$ is independent of $T$, so is just incorporated into the measure. 

The above arguments then show that $S[h^*[T],T]$ can be written as a boundary integral. This is the \em same \em as we would get if we take the boundary action (\ref{23c}) and do the gaussian integration over the boundary values, however, with the constraint (\ref{2.19c}). This means that we should not integrate over all fields $\a_i$ on the boundary, but only those which admit a curl-free extension into the interior. In particular
\be\label{2.27c}
\oint \a_kds_k=0\,.
\ee

Although the derivation of (\ref{2.7e}) given above is simple, it is instructive to consider the problem from the point of view of 2d metrics. 
In two dimensions any infinitesimal change in the metric has a decomposition
\be\label{10}
h_{ij}=\p_i\a_j+\p_j\a_i+g_{ij}\Phi\,,
\ee
where $\a_i(x)$ is a vector field and $\Phi$ is a scalar field. This is equivalent to the well-known fact that all 2d metrics are locally conformally flat: that is there is a coordinate system $\{x'\}$ is which $ds^2=e^\Phi dx'_idx'_i$.
The first two terms in (\ref{10}) correspond to an infinitesimal diffeomorphism $x_i\to x_i'=x_i+\a_i(x)$ of the euclidean metric, and $\Phi\sim e^\Phi-1$ is the infinitesimal change in the conformal factor. This is not completely unique: if $\Phi\to\Phi+\Phi'$ where $\p_k\p_k\Phi'=0$, the change may be absorbed into a shift in $\a$, because only the curvature  $\p_k\p_k\Phi$ is an invariant property of the manifold in two dimensions. 

Substituting the decomposition (\ref{10}), conservation then implies that
\be\label{2.10c}
\p^iT_{ij}\propto\p^i\p_i\a_j+\p^i\p_j\a_i-2\p_j\p^k\a_k-\p_j\Phi=0\,,
\ee
so the scalar curvature vanishes:
\be
\p^j\p_j\Phi=\p^i\p_i\p^j\a_j-\p^j\p_j\p^i\a_i=0\,,
\ee
$\Phi$ can be absorbed into a redefinition of $\a_i$. This is more easily seen in complex coordinates: from (\ref{2.10c})
\be
\p_z\Phi=4\p_z\p_\zb\a_z-2\p_z(\p_\zb\a_z+\p_z\a_\zb)=2\p_z\p_\zb\a_z-2\p_z\p_z\a_\zb\,,
\ee
and
\be
\p_\zb\Phi=2\p_\zb\p_z\a_\zb-2\p_\zb\p_\zb\a_z\,.
\ee
Thus we see that 
$\square\Phi=2\p_\zb\p_z\Phi+2\p_z\p_\zb\Phi=0$, so the curvature vanishes as claimed, and we can write
\be
\Phi=2f(z)+2\bar f(\zb)\,,
\ee
where
\be
\p_z(\p_\zb\a_z-\p_z\a_\zb)=f'(z)\,,\quad \p_\zb(\p_\zb\a_z-\p_z\a_\zb)=-\bar f'(\zb)\,,
\ee
so we may take
\be
\p_\zb\a_z-\p_z\a_\zb=f(z)-\bar f(\zb)\,.
\ee
If we then define (in a simply connected region with some fixed interior point $z_0$)
\be
\tilde\a_z=\a_z+\int_{\zb_0}^{\zb}\bar f(\zb')d\zb'\,,\quad \tilde\a_\zb=\a_\zb+\int_{z_0}^z f(z')dz'\,,
\ee
we have 
\be\label{2.18c}
h_{z\zb}=\a_{z,\zb}+\a_{z,\zb}+(1/2)\Phi=\a_{z,\zb}+\bar f(\zb)+\a_{z,\zb}+f(z)=\tilde\a_{z,\zb}+\tilde\a_{z,\zb}\,,
\ee
and
\be\label{2.19c}
\p_\zb\tilde\a_z-\p_z\tilde\a_\zb=0\,.
\ee
Therefore, at the saddle point, the metric is flat and moreover we may take $\Phi=0$ and $\tilde\a_{i,j}=\tilde\a_{j,i}$ (also in Cartesian coordinates). This means that we could parametrize $\tilde\a_i=\p_i\phi$ where $\phi$ is a scalar potential, but in general it is more convenient not to do so. From here on we drop the tilde over $\a$.

To summarize, the entire action on the right hand side of (\ref{9}) is equivalent to a boundary action
(\ref{23c}) or (\ref{2.23d}), with the constraint (\ref{2.27c}).
The fact that the effective action for $\alpha$ is a total derivative is, from this point of view, at the heart of the property that the $T\overline T$ deformation is `solvable', even when the undeformed theory is not integrable. As we shall show, many of the known results follow from this, as well as new ones in situations where arguments based on translational invariance fail.

\section{Cylinder and torus}\label{sec:cyl}
Although the local action for $\alpha$ is a total derivative, since only $h$, not $\alpha$, needs to be single-valued, there can be a non-trivial contribution when the topology allows for $\a$ to increase by a constant amount around a cycle. As an example consider the flat torus formed by identifying opposite edges of a parallelogram with vertices at 
$0,L,L',L+L'$ where $L=L_1+iL_2$, $L'=L'_1+iL'_2$.

We may  integrate the action for $\a$ out the boundary of the parallelogram, giving a contribution from each edge proportional to 
$\int\e_{jl}\a_j\p_k\a_lds_k$,
where $ds_k$ is the line element. For clarity suppose the edge is parallel to the $x_1$ axis, so we have
\be\label{13}
\int(\a_1^+\p_1\a_2^+-\a_2^+\p_1\a_1^+)dx_1-\int(\a_1^-\p_1\a_2^--\a_2^-\p_1\a_1^-)dx_1\,,
\ee
where $\pm$ are the values on the upper and lower boundaries respectively. Since $\p_1\a_k$ is continuous, if we write $\a_k^+-\a_k^-=[\a_k]_2$, we see that $\p_1[\a_k]_2=0$, so that (\ref{13}) simplifies to
\be
[\a_1]_2\int\p_1\a_2dx_1-[\a_2]_2\int\p_1\a_1dx_1=[\a_1]_2[\a_2]_1-[\a_2]_2[\a_1]_1\,,
\ee
with a similar contribution from the other pair of edges.  

The fact only the discontinuities in $\a$ contribute to the action, and that they are constant along each edge, shows that in fact only \em constant \em flat metrics $h_{ij}$ contribute. It is then simpler to go back to (\ref{9}) and write
\be
e^{\delta t{\cal O}}= (\delta t/A)^{-1}\int\prod_{i,j=1}^2dh_{ij}
e^{-(A/4\delta t)(h_{11}h_{22}-h_{12}^2)+Ah_{ij}T_{ij}}\,,
\ee
where we have incorporated some inessential factors into the measure, and $A$ is the total area of the torus.

Thus if 
$F^{(t)}(\{g_{ij}\})=-\log Z^{(t)}(\{g_{ij}\})$ is the reduced free energy, we have
\be\label{eq:dh}
    e^{-F^{(t+\delta t)}(\{\delta_{ij}\})}=\frac{A}{\delta t}\int\prod_{i,j}dh_{ij}\,
e^{-(A/4\delta t)(h_{11}h_{22}-h_{12}^2)-F^{(t)}(\{\delta_{ij}+h_{ij}\})}\,.
\ee

For $\delta t$ small, we can use the method of steepest descent, expanding $F^{(t)}$ in powers of $h$. 
The calculation is straightforward, but in order to understand the form of the result, 
it is useful to consider the following general integral which will serve as a template for all subsequent similar calculations.

\subsubsection{A small lemma}\label{3.0.1}
Suppose we have an integral over $N$ variables $\{X_j\}$ of the form
\be
e^{-F^{(t+\delta t)}(\{X\})}=(4\pi\delta t)^{-N/2}(\det M)^{1/2}\int \prod_jdx_j\,e^{-(1/4\delta t)\sum_{ij}x_iM_{ij}x_j-F^{(t)}(\{X+x\})}\,,
\ee
where, if the matrix $M$ is not positive definite, we suitably deform the contours to make the integral well-defined.
For small $\delta t$ we may use the method of steepest descents, expanding $F{(t)}$ so that the exponent is
\be
-F^{(t)}(\{X\})-(1/4\delta t)\sum_{ij}x_iM_{ij}x_j-\sum_jx_j\p_{X_j}F^{(t)}-\ffrac12\sum_{i,j}x_ix_j\p_{X_i}\p_{X_j}F^{(t)}+\cdots
\ee
The saddle is at $x_i=-2\delta t\sum_jM^{-1}_{ij}\p_{X_j}F^{(t)}$, at which point the expression above becomes
\be
-F^{(t)}(\{X\})+\delta t\sum_{ij}M^{-1}_{ij}(\p_{X_i}F^{(t)})(\p_{X_j}F^{(t)})+O(\delta t^2)\,.
\ee
However, the fluctuation matrix is now shifted according to
\be
(1/4\delta t)M_{ij}\to (1/4\delta t)M_{ij}+\ffrac12\p_{X_i}\p_{X_j}F^{(t)}=(1/4\delta t)M_{ij}+\ffrac12H_{ij}\,,
\ee
so that the prefactor is modified according to
\begin{eqnarray}
&&(\det M)^{1/2}\to (\det M)^{1/2}\det\left(M+2\delta t\,H)\right)^{-1/2}\\
&&=(\det\big(1+2\delta t\,M^{-1}H)\big)^{-1/2}
=e^{-\delta t\,{\rm Tr}(M^{-1}H)+O(\delta t^2)}\,.
\end{eqnarray}
Thus
\be\label{24}
\p_tF^{(t)}=-\sum_{ij}M^{-1}_{ij}(\p_{X_i}F^{(t)})(\p_{X_j}F^{(t)})
+\sum_{ij}M^{-1}_{ij}(\p_{X_i}\p_{X_j}F^{(t)})\,.
\ee
Note that same matrix $M^{-1}$ occurs in both terms. The equation for the partition function 
$Z^{(t)}=e^{-F^{(t)}}$ is even simpler:
\be\label{25}
\p_tZ^{(t)}=\sum_{ij}M^{-1}_{ij}(\p_{X_i}\p_{X_j}Z^{(t)})\,.
\ee
Of course this could have been found directly by writing
\be\label{26}
Z^{(t+\delta t)}(\{X\})=(4\pi\delta t)^{-N/2}(\det M)^{1/2}\int \prod_jdx_j\,e^{-(1/4\delta t)\sum_{ij}x_iM_{ij}x_j}
Z^{(t)}(\{X+x\})\,,
\ee
and Taylor expanding the last factor.

\begin{center}
\rule{5cm}{0.5pt}
\end{center}

Going back to the torus we have
\be\label{3.14a}
\p_t F=2A^{-1} \e_{ik}\e_{jl}\left(\frac{\p^2F}{\p h_{ij}\p h_{kl}}-\frac{\p F}{\p h_{ij}}\frac{\p F}{\p h_{kl}}\right)\,,
\ee
or, for the partition function
\be\label{3.15a}
\p_t Z=2A^{-1} \e_{ik}\e_{jl}\frac{\p^2Z}{\p h_{ij}\p h_{kl}}\,,
\ee

Computing the second variation of $F$ or $Z$ with respect to the metric turns out to be rather subtle. There is explicit dependence through the lengths $(L,L')$ which 
shift according to $L\to L+\delta L$, where
\be
\delta L_1=\ffrac12h_{11}L_1+\ffrac12h_{12}L_2\,,\quad \delta L_2=\ffrac12h_{21}L_1+\ffrac12h_{22}L_2\,,
\ee
and similarly for $L'$. However this suffices to give only the first-order variation.

It is clearer to proceed by realizing that the first-order response of $F$ to a uniform shift in the metric is given by the integral of the stress tensor: thus
\be
\p_{h_{kl}}F=\int\langle T_{kl}(x)\rangle d^2x=(1/2)(L_k\p_{L_l}+L_k'\p_{L_l'})F\,,
\ee
so that, by translational invariance,
\be\label{3.18b}
\langle T_{kl}(0)\rangle =(1/2A)(L_k\p_{L_l}+L_k'\p_{L_l'})F\,.
\ee
In general, for any local operator $\cal O$,
\be
\p_{h_{kl}}\langle{\cal O}(0)\rangle=\int\langle T_{kl}(x){\cal O}(0)\rangle_c d^2x=(1/2)(L_k\p_{L_l}+L_k'\p_{L_l'})\langle{\cal O}(0)\rangle\,.
\ee
Hence
\be\label{3.20d}
\p_{h_{ij}}\p_{h_{kl}}F=\iint\langle T_{ij}(x)T_{kl}(x)\rangle_c d^2xd^2x'
=A\int\langle T_{ij}(x)T_{kl}(0)\rangle_c d^2x
\ee
\be\label{3.20c}
=(A/4)(L_i\p_{L_j}+L_i'\p_{L_j'})(1/A)(L_k\p_{L_l}+L_k'\p_{L_l'})F\,.
\ee
It is the appearance of the factors of $A$ and $(1/A)$ which modify this result from what we would find by expanding $F$ to second order in $h_{ij}$ taking into account only the explicit dependence through $\delta L$ and  $\delta L'$.\footnote{This factor was omitted in an earlier version of the present paper.}

This may also be understood by realizing that (\ref{3.15a}) is really the specialization of a functional derivative to the case where the metric is constant:
\be
\p_tF=2A^{-1}\e_{ik}\e_{jl}\int d^2x\int d^2x'\left.\frac{\delta}{\delta h_{ij}(x)}\right|_{h_{ij}(x)=h_{ij}}
\left(\left.\frac{\delta}{\delta h_{kl}(x')}\right|_{h_{kl}(x')=h_{kl}}\!\!\!\!F\right)\,,
\ee
where 
\be
\left.\frac{\delta}{\delta h_{kl}(x')}\right|_{h_{kl}(x')=h_{kl}}\!\!\!\!F=\frac1A\frac{\p F}{\partial h_{kl}}\,.
\ee

(\ref{3.14a}) then gives
\begin{eqnarray}
\p_tF&=&\ffrac12\e_{ik}\e_{jl}\left[(L_i\p_{L_j}+L_i'\p_{L_j'})(1/A)(L_k\p_{L_l}+L_k'\p_{L_l'})F\right.\nonumber\\
&&\qquad\left.-((L_i\p_{L_j}+L_i'\p_{L_j'})F)((1/A)(L_k\p_{L_l}+L_k'\p_{L_l'})F)\right]\,,
\end{eqnarray}
or, equivalently for the partition function
\be
\p_tZ=\ffrac12\e_{ik}\e_{jl}(L_i\p_{L_j}+L_i'\p_{L_j'})A^{-1}(L_k\p_{L_l}+L_k'\p_{L_l'})Z
\ee
\be
\!\!\!\!=\left[(L_1\p_{L_1}+L_1'\p_{L_1'})A^{-1}(L_2\p_{L_2}+L_2'\p_{L_2'})
-(L_1\p_{L_2}+L_1'\p_{L_2'})A^{-1}(L_2\p_{L_1}+L_2'\p_{L_1'})\right]Z\,.
\label{3.23b}
\ee
Now write $Z=A\,\mathfrak Z$ where $A=L_1L_2'-L_2L_1'$ and use
\be
(L_1\p_{L_1}+L_1'\p_{L_1'})A=(L_2\p_{L_2}+L_2'\p_{L_2'})A=A\,,\quad (L_1\p_{L_2}+L_1'\p_{L_2'})A=(L_2\p_{L_1}+L_2'\p_{L_1'})A=0\,.
\ee
This gives
\begin{eqnarray}
\p_t\,\mathfrak Z&=&(1/A)\left[(L_1\p_{L_1}+L_1'\p_{L_1'})(L_2\p_{L_2}+L_2'\p_{L_2'})
-(L_1\p_{L_2}+L_1'\p_{L_2'})(L_2\p_{L_1}+L_2'\p_{L_1'})\right.\nonumber\\
&&\qquad\qquad\left.+(L_1\p_{L_1}+L_1'\p_{L_1'})\right]\mathfrak Z\,.
\end{eqnarray}
The last term is now canceled by the terms we get moving the derivatives to the right in the second term, so finally
\be\label{3.26b}
\p_t\,\mathfrak Z=(1/A)(L_1L_2'-L_2L_1')(\p_{L_1}\p_{L_2'}-\p_{L_2}\p_{L_1'})\,\mathfrak Z
=(\p_L\wedge\p_{L'})\,\mathfrak Z\,.
\ee

We also record the equation satisfied by $Z$:
\be\label{3.27b}
\p_tZ
=(\p_{L_1}\p_{L_2'}-\p_{L_2}\p_{L_1'})Z
-(1/A)[(L_2\p_{L_2}+L_2'\p_{L_2'})+(L_1\p_{L_1}+L_1'\p_{L_1'})]Z\,,
\ee
which may also be written, using (\ref{3.18b}),
\be
\p_tZ=(\p_L\wedge\p_{L'})Z-\langle\Theta\rangle\,Z\,.
\ee
where $\Theta$ is the trace $T_{jj}$.

 Although (\ref{3.26b}) is a simple PDE with constant coefficients, and therefore soluble by elementary methods,
 it hides some complexity. This is because the torus partition function depends not on four real variables, but only three, typically taken to be the length of one of the edges and the complex dimensionless modular parameter $\tau$.
 In terms of these parameters it is more complicated. 
It is invariant under a common rotation of the vectors $L$ and $L'$, and the deformation differential operator preserves this.

\subsubsection{Alternate derivation}
There is a more heuristic derivation of (\ref{3.27b}) directly from the saddle-point equations 
(\ref{2.8c}, \ref{2.9c}). In Cartesian components, pointwise in $x$,
\be
h_{11}=(16\delta t)\langle T_{22}\rangle\,,\quad h_{22}=(16\delta t)\langle T_{11}\rangle\,,\quad h_{12}=-(16\delta t)\langle T_{12}\rangle\,,
\ee
where, as before,
\be
\langle T_{ij}\rangle= (1/A)(L_i\p_{L_j}+L_i'\p_{L_j'})F
\ee
and the action at the saddle point is just proportional to the second term
\be
\int T_{ij}h_{ij} d^2x=A\e_{ik}\e_{jl}(L_i\p_{L_j}+L_i'\p_{L_j'})(1/A)(L_k\p_{L_l}+L_k'\p_{L_l'})F\,,
\ee
which gives (\ref{3.20c}) directly if less rigorously.

\subsection{Equivalence to Zamolodchikov's method}\label{sec3.1d}

We now discuss how our method compares with that of Ref.~\cite{Zam}. 
To $O(\delta t)$, the correction to the free energy is 
\be
\delta F=2\delta t\int\langle \e_{ik}\e_{jl}T_{ij}(x)T_{kl}(x)\rangle d^2x\,.
\ee
Zamolodchikov \cite{Zam} argued that in fact $(\p/\p y_m)\e_{ik}\e_{jl}T_{ij}(x)T_{kl}(x+y)$ is a total derivative with respect to $x$, and therefore is expectation value vanishes in translational invariant geometries. This is straightforward to show in Cartesian coordinates, using the identity
\be
(\p/\p y_m)\e_{ik}=\e_{mk}(\p/\p y_i)+\e_{im}(\p/\p y_k)\,.
\ee
The derivative in the second term vanishes on $T_{ij}(x)T_{kl}(x+y)$ by conservation, and the first is 
\be
T_{ij}(x)(\p/\p y_i)T_{kl}(x+y)=T_{ij}(x)(\p/\p x_i)T_{kl}(x+y)=(\p/\p x_i)[T_{ij}(x)T_{kl}(x+y)]\,,
\ee
again by conservation. This implies that 
\be\label{33}
\langle \e_{ik}\e_{jl}T_{ij}(x)T_{kl}(x)\rangle=\langle \e_{ik}\e_{jl}T_{ij}(x)T_{kl}(x')\rangle\,,
\ee
for all $x'$, in any translationally invariant geometry, such as the torus. By inserting a complete set of eigenstates 
$|n\rangle$ of the generators $(E,P)$ of translations along and around the cylinder, Zamolodchikov \cite{Zam} argued that this implies
\be
\e_{ik}\e_{jl}\langle n|T_{ij}T_{kl}|n\rangle=\e_{ik}\e_{jl}\langle n|T_{ij}|n\rangle\langle n|T_{kl}|n\rangle\,.
\ee
From this he was able to deduce the deformation equation for the energy eigenvalues
\be\label{35}
\p_tE_n^{(t)}(\ell)=-E_n^{(t)}(\ell)\p_\ell E_n^{(t)}(\ell)-P_n^2/\ell\,,
\ee
where $\ell$ is the circumference of the cylinder. 

However, another way to proceed from (\ref{33}) is to average the right hand side over $x'$, giving 
\be
\p_t F=2A^{-1}\int\int\langle \e_{ik}\e_{jl}T_{ij}(x)T_{kl}(x')\rangle d^2xd^2x'\,,
\ee
where $A$ is the area of the torus. We recognize $\int T_{ij}(x)d^2x$ as the response of the free energy to a uniform change in the metric, so that
\be\label{2.13}
\p_t F=2A^{-1} \e_{ik}\e_{jl}\left(\frac{\p^2F}{\p h_{ij}\p h_{kl}}-\frac{\p F}{\p h_{ij}}\frac{\p F}{\p h_{kl}}\right)\,,
\ee
or, for the partition function
\be\label{3.41d}
\p_t Z=2A^{-1} \e_{ik}\e_{jl}\frac{\p^2Z}{\p h_{ij}\p h_{kl}}\,,
\ee
which is equivalent to our (\ref{3.14a}, \ref{3.15a}) above. 

Despite the formal equivalence of the two approaches, it is in fact quite laborious to deduce Eq.~(\ref{35})  for the deformation of the energy eigenvalues from (\ref{3.27b}). This we now show.

The torus partition function may be written in a standard way by quantizing on a circle passing through the vertices at $0$ and $L$. The circumference is $|L|=\sqrt{L_1^2+L_2^2}$, and the periodic imaginary time is $A/|L|$.  
An individual term in the eigenstate expansion of $Z$ is then
\be\label{3.37b}
z=e^{-(A/|L|)E(|L|,t)+i(P/|L|)(L_1L_1'+L_2L_2')}=e^{-(A/|L|)E(|L|,t)+i(k/|L|^2)(L_1L_1'+L_2L_2')}\,,
\ee
where $E(|L|,t)$ and $P=k/|L|$ ($k\in2\pi{\bf Z}$) are the eigenvalues of the hamiltonian and momentum respectively.

Without loss of generality we can orient the parallelogram so that $L_2\ll L_1$. Since the differential operator in (\ref{3.27b}) is first order in 
$\p_{L_2}$, we need keep only terms up to $O(L_2)$, whence (\ref{3.37b}) becomes
\be
z=e^{-(L_2'-(L_2L_1'/L_1))E(L_1,t)+i(k/L_1^2)(L_1L_1'+L_2L_2')}\,.
\ee

Then, setting $L_2=0$ after differentiating,
\begin{eqnarray}
\p_{L_1}\p_{L_2'}z&=&(-\p_{L_1}E(L_1,t))\,z+(-E(L_1,t))(-L_2'\p_{L_1}E(L_1,t)-ik(L_1'/L_1^2))\,z\,,\\
\p_{L_2}\p_{L_1'}z&=&((1/L_1)E(L_1,t))\,z+(ik/L_1)((L_1'/L_1)E(L_1,t)+ik(L_2'/L_1^2))\,z\,,
\end{eqnarray}
so that
\be\label{3.41b}
(\p_{L_1}\p_{L_2'}-\p_{L_2}\p_{L_1'})z=(L_2'E\p_{L_1}E+k^2(L_2'/L_1^3))\,z
+(-\p_{L_1}E-(1/L_1)E)\,z\,.
\ee
The remaining terms in (\ref{3.37b}) are 
$$
-(1/L_1L_2')(L_1\p_{L_1}+L_1'\p_{L_1'})z
$$
\be
=-(1/L_1L_2')(-L_1L_2'\p_{L_1}E-ikL_1'/L_1)z-(1/L_1L_2')(+ikL_1'/L_1)z=\p_{L_1}E\,z\,,
\ee
and
\be
-(1/L_1L_2')(L_2\p_{L_2}+L_2'\p_{L_2'})z=-(1/L_1L_2')(-L_2'E)z=(1/L_1)E\,z
\ee
These cancel the last two terms in (\ref{3.41b}).

We should now equate the result with $\p_t z\sim -L_2'\p_tE\,z$.
This yields Zamolodchikov's equation (\ref{35}). 
Equivalently, we can apply (\ref{3.14a}) to $-\log z$, with the same conclusion.

\subsection{Finite cylinder}\label{sec3.2}

A similar analysis may be applied to an $L\times L'$ cylinder with open boundaries along $x_2=0,L'$. In that case, the cycle around $x_1$ allows $\a_1$ to have a discontinuity, but this is no longer the case along $x_2$. Instead, writing
$\a_1(x_1)=h_{11}x_1+\tilde\a_1(x_1)$, where $\tilde\a_1(x_1)$ is periodic modulo $L$, the total derivative integrates up to boundary term
\be\label{3.44b}
\int_0^{L_1}(\a_2\p_1\tilde\a_1-\tilde\a_1\p_1\a_2)\,dx_1\,.
\ee

If the boundary is invariant under reparametrizations of $x_1$, the boundary free energy is independent of $\tilde\a_1$, so integrating over this implies that $\p_1\a_2=0$. Therefore the integration over $h_{ij}$ is concentrated on constant metric
deformations $h_{11}$ and $h_{22}$, the main difference from the torus being that now $h_{12}=0$. We could have found the same result by assuming that the integration is over only constant metrics $h_{ij}$, but that $h_{12}$ decouples if we assume the conformal boundary condition $T_{12}=0$.\footnote{Note that, despite its name, this boundary condition also makes sense in the non-conformal case. In space-time it corresponds to zero momentum flux across the boundary.} 

Thus, comparing with (\ref{3.14a}), we have
\be\label{3.45b}
\p_tF=A^{-1}\frac{\p^2F}{\p h_{11}\p h_{22}}=A^{-1}\iint\langle T_{11}(x)T_{22}(x')\rangle_c d^2xd^2x'\,.
\ee
The calculation is similar to that of the torus except that we have to be careful about assuming translational invariance. Reflection symmetry under $x_1\to-x_1$ implies that $\langle T_{12}(x)\rangle=\langle T_{21}(x)\rangle=0$, and hence conservation implies that
$\p_{x_2}\langle T_{22}(x_1,x_2)\rangle=0$. Together with translational symmetry in $x_1$  this means that 
$\langle T_{22}(x)\rangle$ is constant, and so we may write
\be
\langle T_{22}(0)\rangle=A^{-1}\int \langle T_{22}(x)\rangle d^2x=(1/LL')L'\p_{L'}F\,,
\ee
and therefore, from (\ref{3.45b}),
\be
\p_tF=L\p_L[(1/LL')L'\p_{L'}F]\,.
\ee
Thus, for the partition function,
\be\label{3.48c}
\p_tZ=L\p_L[(1/L)\p_{L'}Z]=\big(\p_L-(1/L)\big)\p_{L'}Z
\ee

Note that this equation is asymmetrical between $L$ and $L'$, as is allowed by the geometry. The second term can  be simply removed by setting $Z=L\,\mathfrak Z$. 

If we write $Z$ as a sum over eigenstates of translations along $x_1$, it is a trace, and each term gives
\be
\p_te^{-LE(L',t)}=-L(\p_tE)e^{-LE(L',t)}=\big(\p_L-(1/L)\big)\p_{L'}e^{-LE(L',t)}=(E\p_{L'}E)e^{-LE(L',t)}\,,
\ee
giving Zamoldochikov's equation (\ref{35}) with $P$ set equal to zero. 

On the other hand, writing $Z$ as a sum over eigenstates of translations along $x_2$, it is a sum of terms of the form
$f(L,t)e^{-L'E(L,t)}$ where $f$ is a projection (squared) onto the boundary state. 
After some algebra we find that $E$ again satisfies
(\ref{35}) with $P=0$, and furthermore
\be\label{3.50b}
\p_tf=-\big(\p_L-(1/L)\big)(fE)\,.
\ee
Once again, $f/L$ satisfies a simpler equation.

When $L'\gg L$, the leading term in the partition function has the form
\be
Z\sim e^{s_a+s_b-L'E_0}\,,
\ee
where now $E_0$ is the ground state energy. For a CFT, $s_{a,b}$ are universal constants depending only on the (conformal) boundary conditions, the Affleck-Ludwig boundary entropies \cite{Aff}. They each satisfy
\be
\p_ts=-E_0\p_Ls-(\p_L-(1/L))E_0\,.
\ee
Since at $t=0$ the $s_{a,b}$ are independent of $L$, it follows that $s_{a,b}(t)-s_{a,b}(0)$ is a universal function, depending only on $c$ but not the type of boundary condition.

We note that (\ref{35}) with $P$ set to zero was shown to be true for the case of free fermions and conjectured to be more generally valid in \cite{Cav}. They also derived a result equivalent to (\ref{3.50b}). However we stress that this is in fact the case only for conformal boundary conditions. In principle the boundary free energy could depend on $(\p_1\tilde\a_1)^2$, for example. 

If we assume that $Z^{(t)}$ has the leading form $e^{-(f_tLL'+2\sigma_tL+\cdots)}$, then (\ref{35}) implies that
$\p_tf_t=-f_t^2$ as before, but also that $\p_t\sigma_t=-f_t\sigma_t$. This shows that in a deformed CFT, when we can consistently set $f_t=0$, the surface tension $\sigma_t$ does not evolve. 

\subsection{Solution for the partition function.}\label{sec3.3d}

We now discuss the solution of the PDE for the partition function, when $Z^{(0)}$ corresponds to a CFT.

It is simpler to consider the case of the finite $L\times L'$ cylinder, (\ref{3.48c}), which when expressed  
in terms of $\vZ=Z/L$ satisfies a linear PDE with constant coefficients
\be\label{3.60e}
\p_t\vZ=\p_L\p_{L'}\vZ\,.
\ee
Let us first examine the expectations based on the solutions of
(\ref{35}). As discussed in the previous section, if we require that $Z^{(t)}(L,L')$ may be expanded as a sum of terms of the form 
$e^{-LE_n^{(t)}(L')}$, or, alternatively $f_n^{(t)}(L)e^{-L'E_n^{(t)}(L)}$, then in both cases $E_n^{(t)}$ satisfies (\ref{35}) with $P=0$.
The initial condition is $E_n^{(0)}(L)=\pi\Delta_n/L$, where in the first case $\Delta_n=x_n^b-c/24$ where $x_n^b$ is a boundary conformal weight, and in the second case $\Delta_n=2x_n-c/6$ where $x_n$ is a bulk conformal weight. The solution is \cite{Zam,Smi}
\be
E_n^{(t)}(L)=\frac L{2t}\left(1-\sqrt{1-\frac{4\pi\Delta_n t}{L^2}}\right)\,.
\ee
This becomes singular at $t=L^2/4\pi\Delta_n$. Since in general there are only a finite number of weights with $\Delta_n<0$ (including the ground state), while there are an infinite number with $\Delta_n>0$, the expansion behaves differently for $t<0$ and $>0$. For $t<0$, if we assume for simplicity that only the ground state has $\Delta<0$, this implies a singularity in the partition function at $L=O((ct)^{1/2}$, which is usually identified as a Hagedorn transition \cite{Cas}. Another way to see this is to note that for $\Delta_n\gg L^2/t$, the spectrum behaves like $E_n^{(t)}\sim(\pi\Delta_n/t)^{1/2}$ so that 
\be
Z^{(t)}\sim \int\rho(\Delta)e^{-L'(\pi\Delta/t)^{1/2}}d\Delta\,,
\ee
where the density of states $\rho\sim e^{{\rm cst.}(c\Delta)^{1/2}}$, leading to a divergence at $t\sim L^2/c$.

For $t>0$ the situation is different. As soon as $t\not=0$  there is an infinite number of $\Delta_n$ for which $E^{(t)}_n$ is complex, with a real part $L/2t$, independent of $\Delta_n$, and therefore each giving a contribution of the same order to the sum over $n$.
We now argue that the PDE (\ref{3.48c}), or equivalently (\ref{3.60e}), with CFT initial conditions, has in fact a well-defined solution for all $t$, but that for $t>0$ it does not have a convergent expansion in terms of the form $e^{-LE_n^{(t)}(L')}$ or $f_n^{(t)}(L)e^{-L'E_n^{(t)}(L)}$.

Eq.~(\ref{3.60e}) is to be solved in the domain $L,L'>0$. Since $Z^{(t)}$ behaves as $L'\to\infty$ like
$e^{-L'E_0^{(t)}(L)}$, it is natural to try to solve by Laplace transform, defining 
\be
\widetilde\vZ^{(t)}(L;s)=\int_0^\infty e^{-sL'}\vZ^{(t)}(L,L')dL'\,.
\ee
However, this makes sense only if it converges as $L'\to0$, but in fact $Z^{(0)}\sim e^{(\pi c/6)(L/L')}$ in that limit.
Let us therefore subtract off this term for all $t$ and consider
\be
\overline Z^{(t)}(L,L')\equiv Z^{(t)}(L,L')-e^{-LE_0^{(t)}(L')}\,,
\ee
where $E_0^{(t)}(L')$ satisfies (\ref{35}), so that the subtraction, and  therefore $\overline Z^{(t)}$, both satisfy the PDE 
(\ref{3.48c}). (If there are more states with $E^{(t)}(L')\leq0$ we should also subtract off their contribution.) 

The subtracted $\overline\vZ$ then vanishes as $L'\to0$, so that (\ref{3.60e}) is then transformed to
\be
\p_t\widetilde\vZ=s\,\p_{L}\widetilde\vZ\,,
\ee
with the solution
\be
\widetilde\vZ^{(t)}(L;s)=\widetilde\vZ^{(0)}(L+st;s)\,,
\ee
where
\be
\widetilde\vZ^{(0)}(L;s)=(1/L)\int_0^\infty e^{-sL'}\left(\sum_ne^{-L'E_n^{(0)}(L)}-e^{(\pi c/6)(L/L')}\right)dL'\,.
\ee
The expression in parentheses behaves as $e^{-2\pi\Delta(L/L')}$ as $L'\to0$, with $\Delta>0$, and so the integral converges absolutely and uniformly and defines
an analytic function of $s$ for sufficiently large ${\rm Re}\,s$. The second term cancels the divergence in the sum over $n$ as $L'\to0$. There is a simple pole at $s=-E_0^{(0)}(L)=(\pi c/24 L)$. The second term then gives a  singularity at $s=0$, which may be estimated by steepest descents, to be of the form $e^{-{\rm cst.}(-sL)^{1/2}}$. Although the first term in principle gives further poles at $s=-\pi\Delta_n/L<0$, in fact the contour cannot be moved beyond this square root  branch point to pick up their contributions. 

The solution for the subtracted $\vZ^{(t)}$ is then
\be
\overline\vZ^{(t)}(L,L')=\int_C\widetilde\vZ^{(0)}(L+st;s)e^{sL'}ds/(2\pi i)\,,
\ee
where as usual $C$ runs parallel to the imaginary axis to the right of all the singularities of the integrand. A pole at
$s=-\pi\Delta/L$ now contributes a term
\be
\frac1{s+\frac{\pi\Delta}{L+st}}
\ee
to the integrand, corresponding to two poles at 
\be
s=-\frac L{2t}\left(1\pm\sqrt{1-\frac{4\pi\Delta t}{L^2}}\right)\,,
\ee
where the lower sign corresponds to the solution $s=-E^{(t)}$ of (\ref{35}) with $P=0$. The branch point now has the form 
$e^{-{\rm cst.}(-s(L+st))^{1/2}}$.

We therefore see that, for $t>0$: 
\begin{itemize}
\item  the pole corresponding to the ground state, with $\Delta=-c/12$ and the lower sign, remains real and positive, approaching, as
$t\to+\infty$, a value $O(t^{-1/2})$;
\item the branch cut now runs from $s=0$ to $s=-L/t$;
\item  the excited states with $\Delta>0$ become complex at some value of $t=O(L^2/\Delta)>0$, with a real part 
$-L/2t$, but are sub-leading for large $L'$ relative to the branch point at $s=0$. 
\end{itemize}

We conclude that for $t>0$ there is in fact a regular solution to (\ref{3.48c}), with an asymptotic expansion for large $L'$
\be
Z^{(t)}(L,L')=f_0^{(t)}(L)e^{-L'E_0^{(t)}(L)}+e^{-LE_0^{(t)}(L')}+\cdots\,,
\ee
where the remaining terms are subleading at large $L'$. The second term does not have an expansion in powers of $e^{-L'}$. 

This also holds at large $L$, with different correction terms.  We expect the first two terms to become exact at large $c$. 
Although the analysis is more complicated for the case of the torus, since $L$ and $L'$ become complex, we expect a similar result to hold. 

This result suggests that the thermodynamics of the system, where we take $L'\gg L=\beta$, is well-defined for $t>0$, despite the partition function not having a spectral decomposition. The free energy per unit length is given by
\be\label{3.72e}
\beta F(\beta)=E_0^{(t)}(\beta)=\frac{\beta}{2t}\left(1-\sqrt{1+\frac{2\pi ct}{3\beta^2}}\right)\,,
\ee
which is regular at all temperatures, unlike the case $t<0$. However the internal energy
$\p_\beta(\beta F)$ approaches a constant $(1/2t)$ at infinite temperature, and is regular at $\beta=0$. This suggests the possibility of a negative temperature phase in which the free energy would be given by the other branch of the square root in
(\ref{3.72e}), analogous to what happens in a quantum spin system.

\section{Domains with boundaries}\label{sec:bound}

\subsection{Disk}\label{sec4.1}
In this section we consider examples where the domain $\cal D$ has a boundary $\p\cal D$.  We first consider the simplest case when $\cal D$ is the disk $|x|\leq R$.

In general the boundary action takes the form (\ref{2.23d})
\be\label{4.1c}
(1/2\delta t)\int_{\p\cal D}[\e_{ij}(\a_i\p_k\a_j)+\Psi\a_k]ds_k-2\int_{\p\cal D}\e_{jk}\a_iT_{ij}ds_k\,,
\ee
where $ds_k$ is a tangential line element, the degrees of freedom are $\a_i(s)$ on the boundary, and the global lagrange multiplier $\Psi$  implements the constraint $\oint\a_kds_k=0$. Note that $ds_k\p_k$ is just the tangential derivative. As long as we continue to write $\a_i$ in cartesian coordinates there are no complications from transporting this vector around the curve and we may use ordinary partial derivatives. 
However, for the disk polar coordinates are most simple, and we should be careful in writing such derivatives of  $(\a_r,\a_\theta)$. The action is\footnote{We adopt the convention that $\a_\theta$ has the dimensions of length. Thus the infinitesimal change in $\theta$ is $(1/r)\a_\theta$.}
\be\label{4.6c}
(1/2\delta t)\int[\a_r(\p_\theta\a_\theta+\a_r)-\a_\theta(\p_\theta\a_r-\a_\theta)+R\Psi\a_\theta]d\theta
-2\int(\a_rT^{rr}+\a_\theta T^{r\theta})Rd\theta\,.
\ee
This simplifies if we assume, as before, the conformal boundary condition $T^{r\theta}=0$, for then we can integrate over $\a_\theta$ explicitly. The first term, however first requires an integration by parts, and this will give an extra contribution if $\a_\theta$ has non-trivial winding around the boundary, as is allowed by the continuity of the metric. This would correspond to deforming the disk into a cone. However, we now argue that this is inconsistent with the saddle point equations and conservation.

In polar coordinates the conservation equations read
\begin{eqnarray}
\p_r(rT^{rr})+\p_\theta T^{r\theta}&=&T^{\theta\theta}\,,\\
\p_r(rT^{r\theta})+\p_\theta T^{\theta\theta}&=&-T^{r\theta}\,,
\end{eqnarray}
and therefore at the saddle point $h_{ij}=h_{ij}^*\propto\e_{ik}\e_{jl}T^{kl}$
\begin{eqnarray}
\p_r(rh_{\theta\theta})-\p_\theta h_{r\theta}&=&h_{rr}\,,\\
-\p_r(rh_{r\theta})+\p_\theta h_{rr}&=&h_{r\theta}\,.
\end{eqnarray}
Integrating the first equation $\int_0^Rdr\int_0^{2\pi}d\theta$ gives
\be
R\int_0^{2\pi}h_{\theta\theta}(R,\theta)d\theta=\int_0^R\int_0^{2\pi}h_{rr}(r,\theta)drd\theta\,.
\ee
However $h_{rr}=2\p_r\a_r$ and $h_{\theta\theta}=(2/r)(\a_r+\p_\theta\a_\theta)$, and therefore
\be
\int_0^{2\pi}\p_\theta\a_\theta(R,\theta)d\theta=0\,.
\ee

Thus $\a_\theta$ is single valued, as claimed.  Varying (\ref{4.6c}) wrt $\a_\theta$
then gives
\be
\a_\theta=\p_\theta\a_r-R\Psi/2\,,
\ee
where in fact $\Psi=0$ to satisfy the constraint  $\oint\a_\theta d\theta=0$.

Since $h_{r\theta}=\p_r\a_\theta+ (1/r)(\p_\theta\a_r-\a_\theta)\propto T^{r\theta}=0$, the condition $\a_\theta=\p_\theta\a_r$ implies that $\p_r\a_\theta$ also vanishes on the boundary. This is also consistent with the curl-free condition, which in polar coordinates reads $\p_r\a_\theta=(1/r)(\p_\theta\a_r-\a_\theta)$. 
Physically, $\a_\theta=\p_\theta\a_r$ means 
that each line element of the boundary moves without changing the direction of its tangent vector:  examples are a uniform dilatation $\a_r=$ constant, or a rigid translation of the disk where $\a_r\sim\cos\theta$, $\a_\theta\sim-\sin\theta$.

On integrating over $\a_\theta$ the action (\ref{4.6c}) then simplifies to
\be\label{4.10d}
(1/2\delta t)\int(-(\p_\theta\a_r)^2+\a_r^2)d\theta-2\int\a_rT^{rr}Rd\theta\,.
\ee
Note the `wrong' sign for the derivative term. Since the original gaussian integral required some rotation of the contours, this is not surprising. 
Integrating over $\a_r$ we  then find 
\be\label{4.14c}
\delta F=-2\delta tR^2\int G(\theta-\theta')\langle T^{rr}(\theta)T^{rr}(\theta')\rangle_cd\theta d\theta'
=\ffrac12\delta tR^2\int G(\theta-\theta')\frac{\delta^2F}{\delta(R\a_r(\theta))\delta(R\a_r(\theta'))}d\theta d\theta'
\,,
\ee
where $G$ is the Greens function for $(\p_\theta^2+1)$:
\be
(\p_\theta^2+1)G(\theta-\theta')=\delta_p(\theta-\theta')\,,
\ee
where $\delta_p$ is the periodic delta function.
In terms of a formal eigenfunction expansion,
\be
G(\theta-\theta')=\frac1{2\pi}\sum_{n\in\mathbb Z}\frac{e^{in(\theta-\theta')}}{1-n^2}\,.
\ee
However, the modes $\a_r\propto e^{\pm i\theta}$ are rigid translations of the disk, for which the free energy does not change, so they may be subtracted off. The modified Greens function  $\widetilde G$ then satisfies
\be
(\p_\theta^2+1){\widetilde G}(\theta)=\delta_p(\theta)-(1/2\pi)(e^{i\theta}+e^{-i\theta})\,,
\ee
and the solution is, for $|\theta|<\pi$,
\be\label{4.18c}
{\widetilde G}(\theta)=(1/2)\big(\text{sign}(\theta)-(\theta/\pi)\big)\sin\theta\,.
\ee
Note that the unphysical  discontinuity at $\theta=\pm\pi$ cancels.

In terms of a mode expansion $\a_r(\theta)=\sum_n\a_ne^{in\theta}$, we can rewrite (\ref{4.14c}) for the partition function
\be
\p_tZ=\frac{R^2}{4\pi}\sum_{n\not=\pm1}\frac1{1-n^2}\frac\p{R\p\a_n}\frac\p{R\p\a_{-n}}Z\,.
\ee
Thus we see that the evolution equation for a disk is more complicated than that for a torus and a finite cylinder, involving all the possible modes of deformation, not only the rotationally symmetric mode. However, we may isolate this by integrating out these modes and considering the partition function at fixed perimeter $2\pi R+\int\a_r(\theta)d\theta=2\pi(R+\a_0)$, that is
\be\label{4.17d}
\widetilde Z(R)\equiv\int Z(\{\a_n\})\prod_{n\not=0}d\a_n\,,
\ee
which satisfies
\be\label{4.18d}
\p_t\widetilde Z=\frac{R^2}{4\pi}\frac1R\p_R\big((1/R)\p_R\widetilde Z\big)=
\frac1{4\pi}\left(\frac\p{\p R}-\frac1R\right)\frac{\p\widetilde Z}{\p R}
\,.
\ee
In writing this we have been careful to keep the factors of $R$ in place, as $R\to R+\a_0$ and $\p_{\a_0}$ acts on this.
The form of this equation may be checked in perturbation theory: in a CFT \cite{Pes}  $Z^{(0)}(R)\propto R^{c/6}$, so according to (\ref{4.14c},\ref{4.18d}) the first order term is
\be
(t/4\pi)(c/6)(c/6-2)R^{c/6-2}=2tR^{c/6}R^2\int\langle T_{rr}(R,\theta)T_{rr}(R,\theta')\rangle d\theta d\theta'\,.
\ee
The $O(c^2)$ term comes from the disconnected part $R^2(\int \langle T_{rr}(R,\theta)\rangle d\theta)^2\propto
(\int\langle\Theta\rangle d^2x)^2$, and the $O(c)$ term from the usual $\langle TT\rangle$ correlator integrated along the boundary. Since the disconnected piece enters only the $n=0$ term in (\ref{4.14c}), this shows that 
in the large $c$ limit (\ref{4.18d}) becomes exact for $Z$, not only $\widetilde Z$.

Although the result (\ref{4.14c}) together with (\ref{4.18c}) is not particularly illuminating, it may be cast into a number of other forms. The most useful is to realize that the local radius of curvature (the inverse of the extrinsic curvature of the boundary) is, to first order in $\a$,
\be
\rho(\theta)=R+(1+\p_\theta^2)\a_r(\theta)=R(1+h_{\theta\theta})\,,
\ee
so that $\delta/\delta\a_r(\theta)=(1+\p_\theta^2)\delta/\delta\rho(\theta)$, and
(\ref{4.14c}) becomes
\be\label{4.20d}
\delta F=\ffrac12\delta t\int\frac\delta{\delta\rho(\theta)}(1+\p_\theta^2)\frac\delta{\delta\rho(\theta)}F d\theta\,,
\ee
or, for the partition function,
\be
\p_tZ=\ffrac12\int\frac\delta{\delta\rho(\theta)}(1+\p_\theta^2)\frac\delta{\delta\rho(\theta)}Z d\theta\,.
\ee

It is also worth noting that from (\ref{4.10d}) the saddle-point equation solution is
\be
\rho(\theta)^*=R+2\delta t R\langle T^{rr}(R,\theta)\rangle=R-2\delta t \frac{\delta F}{\delta\a_r(\theta)}\,,
\ee
or equivalently
\be\label{4.23d}
\p_t\rho=-2(1+\p_\theta^2)\frac{\delta F}{\delta\rho(\theta)}\,.
\ee

\subsection{Solution for the partition function}
Although (\ref{4.18d}) maybe solved by a Greens function, it is more instructive to try a power series solution. This is because in a CFT we can set $f_t=\sigma_t=0$ and then \cite{Pes}, 
\be
\widetilde Z^{(0)}(R)\sim (R/\e)^{c/6}\,,
\ee
where $c$ is the central charge and $\e$ is a UV cut-off. In fact it is useful to consider the more general initial condition 
$\widetilde Z^{(0)}(R)\sim R^{-x}$, where $x=\Delta-c/6$ corresponding to the insertion of a scalar operator of conformal weight
$\Delta$ at the origin, or equivalently a hole of radius $\e\ll R$ with a different conformal boundary condition than that at $r=R$. Note that (\ref{4.18d}) implies that the $\e$ dependence of $\widetilde Z^{(t)}$ is the same as that at $t=0$, so we may drop in the subsequent analysis.  

However if we try a straightforward power series
\be
\widetilde Z^{(t)}(R)=R^{-x}\sum_{n=0}^\infty a_n(t/R^2)^n\,,
\ee
we find
\be
a_n=\frac{(2n-2+x)(2n+x)}{4\pi n}\,a_{n-1}\,,
\ee
so the series is divergent. More generally, we may look for a scaling solution of the form
\be
\widetilde Z^{(t)}(R)=t^{-x/2}Y(u=R^2/t)\,,
\ee
where
\be
(-x/2)Y-uY'=(1/\pi)uY''\,.
\ee
This is related to a confluent hypergeometric equation, and has independent solutions one of which behaves as $u\to\infty$
like $u^{-x/2}$ as expected, but the other as $e^{-\pi u}$. The latter gives an essential singularity as $t\to0+$ which is responsible for the divergence of the naive power series solution. The correct linear combination of these solutions depends on the boundary condition at $R=0$, but it may be shown that the solution with a finite limit as $u\to0$ contains the essential singularity. 
Unfortunately, unlike the case of the torus and cylinder, we do not have an explicit form to check against.

\subsection{General simply connected domain}
Now consider a simply connected domain with a smooth boundary. 
Starting from (\ref{4.1c}), 
the action is exact at the saddle point where
\be
\p_s\a_j=(4\delta t)\e_{ij}\e_{lk}T_{il}\hat t_k-\Psi\e_{jk}\hat t_k\,,
\ee
with a formal solution
\be
\a_j(s)=(4\delta t)\int\e_{ij}\e_{lk}\e(s-s')T_{il}(s')ds'_k-\Psi\e_{jk}s_k\,,
\ee
where $\e(s-s')$ is defined as $+\frac12$ for $0$$<$$s$$-s'$$<$$\ell/2$ and $-\frac12$ for $-\ell/2$$<$$s$$-$$s'$$<$$0$, where $\ell$ is the perimeter length. $\Psi$ is fixed by requiring that $\oint\a_jds_j$: this also eliminates the
unphysical singularity at $s-s'=\ell/2$.
This gives
\be
\a_j(s)=(4\delta t)\int\e_{ij}\e_{lk}G(s-s')T_{il}(s')ds'_k\,,
\ee
where $G(s-s')=\e(s-s')-(s-s')/\ell$. Inserting this into the action and using the conformal boundary condition then gives the generalization of (\ref{4.14c})
\be\label{4.25d}
\delta F=-\ffrac12\delta t\int G(s-s')\langle T^{nn}(s)T^{nn}(s')\rangle_c\,ds\wedge ds'\,,
\ee
where $T^{nn}=T^{kl}\hat n_k\hat n_l$. Note that the sine function in (\ref{4.18c}) is now incorporated in the wedge product $ds\wedge ds'=\e_{ij}ds_ids'_j$.

In fact if we define $\theta=\int ds/\rho(s)$, where $\rho(s)$ is the local radius of curvature, we can see that  (\ref{4.25d})
has precisely the same form as (\ref{4.20d}), although strictly speaking this requires the curve to be convex so that $\theta$ is single-valued. This implies that $\widetilde Z(R)$, defined as the partition function for all (convex) domains of fixed perimeter $2\pi R$, satisfies (\ref{4.18d}). In addition, as it is completely local, (\ref{4.23d}) is valid more generally, with $\theta$ defined as above. 
We will return to this in Sec.~\ref{sec5}.

\subsection{Polygonal domains}

It is instructive to specialize the discussion of the simply connected domain to the case
when the boundary is piecewise linear, \em i.e. \em a polygon.

For each linear segment of the boundary, taken without loss of generality to be the interval $a<x_1<b$ along $x_2=0$, (\ref{2.23d}) gives
\be\label{4.16a}
\int_a^b(\a_2\p_1\a_1-\a_1\p_1\a_2)dx_1=[\a_2\a_1]_a^b-2\int_a^b(\a_1\p_1\a_2)dx_1\,.
\ee
Let $\a_1=\a_1'x_1+\tilde\a_1$, where $\tilde\a_1(a)=\tilde\a_1(b)$. Boundary reparametrization invariance implies that the free energy does not depend on $\tilde\a_1(x_1)$, so we may freely integrate over it, giving the constraint
$\p_1\a_2=0$. The right hand side of (\ref{4.16a}) then becomes  $\a_2[\a_1]_a^b$. 

Thus the measure is concentrated on deformations in which each edge is moved normal to itself by an amount $\a_2$ and stretched by $[\a_1]_a^b$. This is in agreement with the observation in Sec.~\ref{sec4.1} that each line segment moves parallel to itself. In particular this means that the angles between adjacent edges are preserved. 
 If, in complex notation, the edges are  $\ell_je^{i\theta_j}$ with $j=1,\ldots,N$, then each $\ell_j$ may vary subject to
\be\label{4.17a}
\sum_j\ell_je^{i\theta_j}=0\,.
\ee
However, the saddle-point equations restrict this further. Along the above edge, varying with respect to $\a_2$, we have 
\be
[\a_1]=2\delta t\int_a^b\langle T_{22}(x_1)\rangle dx_1=-2\delta t\p_{\a_2}F\,.
\ee
For example, for an $L\times L'$ rectangle, 
\be\label{4.38d}
\delta L= -2\delta t \p_{L'}F\,,\quad \delta L'= -2\delta t \p_{L}F\,.
\ee
These correspond to a uniform deformation of the metric $h_{ij}$, just as for the torus and finite cylinder. The second variation of the free energy may again be expressed in terms of a double integral of the 2-point function of the stress tensor over the domain. However, unlike those cases, there is not enough translational symmetry to write this as a second variation with respect to the edge lengths. Nevertheless (\ref{4.38d}) may be used to give a quick derivation of Zamolodchikov's equation (\ref{35}), see Sec.~\ref{sec5}.

\section{Interpretation as stochastic dynamics}\label{sec5}

In all the examples we have considered so far, the partition function satisfies a linear PDE which is first order in time and second order in the parameters specifying the domain $\cal D$, and is therefore of diffusion type.  The partition function may then be viewed as being proportional to the probability distribution function for a kind of Brownian motion of a particle moving in the parameter space.

There are in fact two different but related processes corresponding to each geometry. These are best illustrated using the example of the lemma of Sec.~\ref{3.0.1}, where (\ref{25}) for the partition function 
\be
\p_tZ^{(t)}=\sum_{ij}M_{ij}^{-1}(\p_{X_i}\p_{X_j}Z^{(t)})
\ee
corresponds to the process\footnote{We use physicists' notation rather the more correct language of stochastic differential equations.}
\be \label{5.2d}
\p_tX^{(t)}_i=\eta_i(t)\,,\quad\mbox{where}\quad\overline{\eta_i(t')\eta_j(t'')}=2M_{ij}^{-1}\delta(t'-t'')\,,
\ee
which follows from expanding in a Taylor series
\be
Z^{t+\delta t}(\{X_i^{(t)}\})=\overline{Z^{(t)}\big(\{X^{(t)}_i+\int_t^{t+\delta t}\eta_i(t')dt'\}\big)}\,.
\ee
Thus
\be
Z^{(t)}(\{X_i^{(0)}\})=\overline{Z^{(0)}(\{X_i^{(t)}\})}\,.
\ee

The second process comes from considering (\ref{24}) for the free energy
\be
\p_tF^{(t)}=-\sum_{ij}M_{ij}^{-1}(\p_{X_i}F^{(t)})(\p_{X_j}F^{(t)})+\sum_{ij}M_{ij}^{-1}(\p_{X_i}\p_{X_j}F^{(t)})\,,
\ee
which corresponds to the process
\be
\p_tX^{(t)}_i=-\sum_jM^{-1}_{ij}\p_{X_j}F^{(t)}+\eta_i(t)\,,
\ee
with the same noise correlations as in (\ref{5.2d}). This version is physically more appealing, as the driving force in the first term corresponds to the saddle point equation, and it looks like standard relaxational dynamics with noise correlations given by the Einstein relation. However the force is given in terms of the instantaneous free energy $F^{(t)}$, not $F^{(0)}$. On the other hand, in the first formulation the actual form of the free energy enters only through the initial conditions.

As the simplest example, consider the finite cylinder treated in Sec.~\ref{sec3.2}, where the equation for the rescaled partition function 
$\vZ^{(t)}=Z^{(t)}/L$ is
\be
\p_t\vZ^{(t)}(L,L')=\p_L\p_{L'}\vZ^{(t)}(L,L")\,,
\ee
which is associated with the Brownian motions
\be\label{Rt}
\p_tL_t=\eta(t)\,,\quad \p_tL_t=\eta'(t)\,,
\ee
where $\overline{\eta(t')\eta'(t'')}=\delta(t'-t'')$, and the other variances vanish. 

Equivalently, in order to avoid the factor of $L$, we may view this is a process in the space of uniform metrics $(h_{11},h_{22})$, whereby the equation for the actual partition function 
\be
\p_tZ^{(t)}(h_{11},h_{22})=4\p_{h_{11}}\p_{h_{22}}Z^{(t)}(h_{11},h_{22})\,,
\ee
corresponds to a similar process to the above. 

For the torus we then have a similar equation
\be
\p_tZ^{(t)}(\{h_{ij}\})=2\e_{ik}\e_{jl}\p_{h_{ij}}\p_{h_{kl}}Z^{(t)}(\{h_{ij}\})\,,
\ee
corresponding to
\be
\p_th_{ij}=\eta_{ij}(t)\,,\quad\mbox{where}\quad\overline{\eta_{ij}(t')\eta_{kl}(t'')}=4\e_{ik}\e_{jl}\delta(t'-t'')\,.
\ee

However in both cases, the noise correlations are not positive semi-definite, which corresponds to the right hand side of the PDEs being hyperbolic rather than elliptic. This casts some doubt on the stochastic interpretation. However, as long as we restrict the domain to be $(L,L">0)$, that is $(h_{11},h_{22}>0)$, for the finite cylinder and $L\wedge L'>0$ for the torus (that is $\det h>0$) the solution for $Z{(t)}$ remains positive on physical grounds. This then requires choosing the correct boundary conditions when $\det h$ vanishes. Since the total mass of the measure on $h_{ij}$ is invariant, this means that $\int_{\det h>0}[dh]Z[h]$ should be conserved. This would then correspond to reflecting boundary conditions on the stochastic process, that is Neumann boundary conditions on the PDE. 

It should also be pointed out that the initial conditions are rather singular. For the finite cylinder and a CFT initial condition,
$Z^{(0)}\sim\exp(Ch_{11}/h_{22})$ for $h_{11}/h_{22}\to\infty$, where the constant $C>0$, with a similar behavior in the opposite limit, so that, viewed as a probability density, $Z^{(0)}$ is not normalizable. Nevertheless, as discussed in Sec.~\ref{sec3.3d}, the solution of the PDE is perfectly well-defined. 

It is also interesting to consider the other stochastic point of view,  based on the free energy. In the same limits we have, initially
\be
\p_th_{11}\sim -\frac{Ch_{11}}{h_{22}^2}+\mbox{noise}\,,\quad \p_th_{22}\sim \frac{C}{h_{22}}+\mbox{noise}\,.
\ee
Thus the ratio $h_{11}/h_{22}$ moves away from either singular limit. In fact for the torus we have the general equation
\be
        \p_th_{ij}\sim\e_{ik}\e_{jl}\p_{h_{kl}}F^{(t)}+\mbox{noise}\,,
\ee
and, by symmetry, $\p_{h_{kl}}F^{(t)}$ vanishes when $L'=iL$. and it make be checked that the hessian is positive definite. Thus the torus becomes more symmetrical, with noisy fluctuations.

In fact the saddle point equations which give the deterministic part of the stochastic process based on the free energy may be used to give a quick derivation of Zamolodchikov's equation (\ref{35}), with $P=0$. This is simpler for the finite cylinder, or the
$L\times L'$ rectangle discussed in Sec.~\ref{sec3.2} for which
\be
\p_t L= -\ffrac12 \p_{L'}F\,,\quad \p_tL'= -\ffrac12\p_{L}F\,.
\ee
If we write this for $F=-\log z=-\log\left(f^{(t)}(L)e^{-L'E{(t)}(L)}\right)$ and impose the condition that the evolution of $E^{(t)}$ should be equivalent to the evolution of $(L,L')$, we find 
\be
L'\p_tE\sim E(L)(\p_t L')L(\p_LE(L))(\p_t L)=-L'E(L)\p_LE(L)\,,
\ee
where we have kept only the leading terms $\propto L'$ (the remaining terms give the evolution equation for $f^{(t)}$ discussed in Sec.~\ref{sec3.2}).

The case of the disk is more interesting. The equation (\ref{4.18d}) corresponds to the attractive Bessel process
\be
\p_tR=-\frac1{4\pi R}+\eta\,,\quad\mbox{where}\quad\overline{\eta(t')\eta(t'')}=(1/2\pi)\d(t'-t'')\,,
\ee
which, for $t>0$, corresponds to a Brownian motion attracted to the origin by a force $\propto 1/R$. If the particle starts at any finite distance from the origin it will hit it with probability one. However, the initial condition $Z^{(0)}\propto R^{c/6}$ corresponds to a constant current at large $R$, so that $Z^{(t)}(R)\sim R^{c/6}$ for $t\ll R^{1/2}$ as expected. The essential singularity of the form $e^{-\pi R^2/t}$ corresponds to rare events when the particle manages to reach the origin on this timescale. 

The case $t<0$ is more interesting as it corresponds to a repulsive Bessel process, in fact a marginal one, where
the deterministic solution $R^2\sim t/2\pi$ exactly balances the Brownian motion $\overline{R^2}\sim t/2\pi$. This process describes the distance from the origin $R$ of a standard 2d Brownian motion, which is known to be recurrent, that is the particle always reaches the origin. 

The second form of the stochastic process in terms of the free energy is given, for a more general domain by (\ref{4.23d}) 
\be\label{5.17d}
\p_t\rho=-2(1+\p_\theta^2)\frac{\delta F^{(t)}}{\delta \rho(\theta)}+\mbox{noise}\,.
\ee

To go further depends on the form of the free energy. There are at least two interesting special cases. The first is when there is a constant surface tension, $F^{(t)}=\int \sigma_t ds=\int\sigma_t\rho d\theta$. We argued earlier that as long as the bulk free energy $f_0$ vanishes (which we can assume in a CFT) then $\sigma_t$ does not evolve. In that case, the deterministic part of (\ref{5.17d}) gives $\rho\sim \mbox{constant}-2\sigma t$. This is to be contrasted with the curvature driven interface dynamics associated with 2d coarsening as described by the Allen-Cahn equation \cite{All,Bra}, in which $\p_t\rho\propto -\delta F/\delta \rho(s)=-\rho^{-1}\delta F/\delta\rho(\theta)$, so that $\p_t\rho\propto-\rho^{-1}$, leading to $\rho\propto t^{1/2}$. 
It is amusing to note that the latter behavior can be found in the present context with $\sigma_t=0$ and a conformal boundary condition,
when $\delta F/\delta\rho(\theta)$ is given by the trace anomaly as $\propto-c/\rho$. The connection between 2d coarsening and CFT has been discussed in \cite{Cug1,Cug2} and it would be interesting to pursue this further.

\section{Summary}
In this paper we have treated the $T\overline T$ (more correctly, the $\det T$) deformation of a local 2d quantum field theory using the path integral approach, decoupling the quadratic term in $T_{ij}$ by a gaussian integral over an auxiliary field $h_{ij}$, which may be viewed as a random metric. For an infinitesimal deformation this is equivalent to an infinitesimal diffeomorphism of flat space, and the action for this is then a total derivative, which then only gives a contribution if there is non-trivial topology, or boundaries. Our method leads to linear PDEs for partition functions which are of diffusion type: first order in the deformation parameter and second order in the linear parameters of the manifold. The solution of these by heat kernel methods yields 
results in agreement with known ones, for example, the evolution of the energy eigenvalues on the cylinder, as long as one evaluates the heat kernel integral in a one-loop approximation. For $t>0$ the sum over such eigenvalues is divergent. Nevertheless we argue that the PDE has a regular solution. While for $t<0$ the deformed theory undergoes a Hagedorn transition at finite temperature, as already noted \cite{Cav}, for $t>0$ there is no transition, but the internal energy is finite and regular at infinite temperature, suggesting the possibility of another thermodynamic branch with negative temperature. 
For the disk the solutions in general have an essential singularities as $t\to0+$. It would be interesting to understand the physical significance of these, if any. 

We pointed out that in all cases the evolution of the partition function may be interpreted as that of the distribution function of a kind of Brownian motion in the moduli space of the domain, in the sense that the deformed partition function in the original domain is given by the expectation value of the partition function in the  domain evolved under the stochastic process. In general the parameters of the domain move towards a more symmetrical configuration as the overall size on average shrinks.
The origin of this behavior is in the form of the local action (\ref{9}) for the infinitesimal metric deformation, which is proportional to
\be
(h_{11}+h_{22})^2-(h_{11}-h_{22})^2-4h_{12}^2\,.
\ee
The first term is a local dilatation and the fact that it comes with a positive sign means that, locally, scales try to execute  a Brownian motion and therefore their fluctuations grow. However, the second and third terms correspond to local shear: they carry the opposite sign and therefore their fluctuations shrink. Of course, this picture is oversimplified since the local fluctuations in $h_{ij}$ are not independent, as evidenced by the fact that the measure is in fact a total derivative.

This behavior may understood (or perhaps not) in an elastic analogy: interpreting 
$h_{ij}=\a_{i,j}+\a_{j,i}$ as the strain tensor, 
the saddle point equation 
\be
(\delta t)T_{ij}\propto \e_{ik}\e_{jl}h^{kl}=g_{ij}h^k_k-h_{ij}
\ee
corresponds to a peculiar stress-strain relation  with an infinite Poisson's ratio, or, equivalently, vanishing bulk modulus and finite Young's modulus. In particular $h_{11}\propto T_{22}$ and $h_{22}\propto T_{11}$: thus stress in the $x$-direction produces extension or contraction in the $y$-direction and \em vice versa\em.

Our methods may also be applied to non-simply connected domains with boundaries, with application, for example, to entanglement entropy. By considering the deformed theory on an annulus with different boundary conditions, we may gain access to information about correlations. 
A further step is to apply this method to the actual correlation functions of deformed local operators. This will be difficult, as our approach implies that it is essential to consider conical deformations of the geometry at the locations of the operators.
 Another interesting extension is to manifolds which do not admit a flat metric, for example $S^2$.
 
 The observant reader may have noticed that, with some modifications, none of the arguments in Sec.~\ref{sec3.1d}, nor in those \cite{Zam}, in fact require the stress tensor to be symmetric. Thus the whole set-up should be capable of generalization to non-Lorentz invariant theories, for example those of Lifshitz type or with Galilean invariance. In fact all that is required is any pair of conserved currents $(J, J')$ and a deformation proportional to $J\wedge J'$. 
  These include the $J\overline T$-deformed theories considered in \cite{Gui}. This will be described in a follow-up paper \cite{CarNR}.

 It is an obvious challenge to generalize the 
the whole set-up to higher dimensions. Some progress has recently been made in \cite{Tay}.
In the Appendix we report the results of an investigation in three dimensions. Another intriguing direction is the apparent analogy of the evolution of the boundary of a simply connected domain with curvature-driven dynamics in 2d coarsening, mentioned at the end of Sec.~\ref{sec5}. Such systems show fractal CFT characteristics on large scales \cite{Cug1,Cug2} while smoothing out on scales $<O(t^{1/2})$. It may be that the $t$-parameter of the $T\overline T$ deformation may be interpretable as real time in this context. 

Since the first version of this paper was announced, a number of other related articles have also been posted, the most relevant to the present work being \cite{Aha,Bon,Dub2,Dat,Aha2}.

\acknowledgments

This work was supported in part through funds from the Simons Foundation. 
The author is particularly grateful to  S.~Dubovsky and V.~Gorbenko for drawing attention to important errors in the first versions of this paper: the factors of the area in Eq.~(\ref{1}), and in the argument that this should be equivalent to (\ref{35}). He also thanks Y.~Jiang and S.~Datta for useful discussions and for pointing out a term in (\ref{4.6c}) which was omitted in the first version,
as B.~le Floch and M.~Mezei for comments on an earlier version.

\appendix\section{Higher dimensions}
In this appendix we examine the obstructions to generalizing the arguments of Sec.~\ref{sec2} to higher dimensions, within certain limitations. We assume flat space (a generalization in AdS space has recently been proposed in \cite{Tay}),
and an action of the form
\be
(1/\delta t)\int f[h] d^dx-\int h_{ij}T^{ij}d^dx\,,
\ee
where $h_{ij}$ is an auxiliary tensor field. We assume that $f[h]$ is scalar and homogeneous in $h$, so that at the saddle point
$h_{ij}=h_{ij}^*$, where $T^{ij}=(1/\delta t)\p_{h_{ij}}\!f[h]$, the total action is proportional to $\int h^*_{ij}T^{ij}d^dx$. 

For this to be topological in the sense discussed in Sec.~\ref{sec2}, it must be expressible as a boundary integral of the form $\int \a^*_i T^{ij}ds_j$. Thus $h^*_{ij}=\a^*_{i,j}$: the action is topological only if $h^*_{ij}$ is a diffeomorphism. For generality, however, we do not assume that $T_{ij}$ and therefore $h_{ij}$ are symmetric (the arguments of Sec.~\ref{sec2} do not in fact depend on this \cite{CarNR}). 

To proceed further, we assume that $f[h]$ is in fact quadratic, so that the fluctuations about the saddle point are independent of $T$. The most general invariant such expression has the form
\be
f[h]=a{\rm Tr}\,h^2+b({\rm Tr}\,h)^2\,,
\ee
where $a$ and $b$ are constants. It is straightforward to check that $h^*_{ij}=\a^*_{i,j}=\a^*_{j,i}$ is a solution of the saddle point equation together with conservation law $T_{ij,j}=0$ only if in fact
\be
f[h]\propto\e_{ikpq\ldots  }\e_{jlpq\ldots}h_{ij}h_{kl}=h_{ii}h_{jj}-h_{ij}h_{ij}\,,
\ee
that is, $a=-b$. This is the natural  generalization of the 2d expression. 

However, it is also necessary to show that the saddle point equation has \em only \em   solutions which are diffeomorphisms. Let us examine the 3d case for simplicity. 
The saddle point equations give (we drop the $^*$ on $h$)
\begin{eqnarray}
T_{11}&\propto& h_{22}+h_{33}\,,\\
T_{12}&\propto&-h_{21}\,,\\
T_{13}&\propto&-h_{31}\,,
\end{eqnarray}
and cyclic permutations, so conservation implies that
\be
h_{22,1}+h_{33,1}=h_{21,2}+h_{31,3}\,,
\ee
and cyclic permutations.
A solution is $h_{ij}=\a_{i,j}$ but it is not the most general. If $h_{ij}$ is a solution so is $h_{ij}+\a_{i,j}$.
We may use this gauge freedom to fix $h_{11}=h_{22}=h_{33}=0$. Then
\begin{eqnarray}
h_{21,2}+h_{31,3}&=&0\,,\\
h_{32,3}+h_{12,1}&=&0\,,\\
h_{13,1}+h_{23,2}&=&0\,.
\end{eqnarray}
So we may write 
\begin{eqnarray}
h_{21}&=&\beta_{1,3}\,,\quad h_{31}=-\beta_{1,2}\,,\\
h_{32}&=&\beta_{2,1}\,,\quad h_{12}=-\beta_{2,3}\,,\\
h_{13}&=&\beta_{3,2}\,,\quad h_{23}=-\beta_{3,1}\,.
\end{eqnarray}
or, equivalently,
\be
h_{ij}= -\e_{ijk} \beta_{j,k}\quad\mbox{(no sum on $j$).}
\ee

This appears to be as far as one may go in this direction. The general solution of the saddle point equation is not a diffeomorphism,
and so the action is not topological. However, more artificially perhaps, one may restrict the integration over $h_{ij}$ to be only over diffeomorphisms, but this does not correspond to deformations which are local when expressed in terms of $T$.
Rather, it leads to integrals over the $\langle TT\rangle$ correlator like those in (\ref{3.20d}) and (\ref{4.14c}).

\end{document}